\documentclass[]{acm_sen_article}

\usepackage{lipsum}
\usepackage{flushend}

\usepackage{graphicx} 
\PassOptionsToPackage{hyphens}{url}
\usepackage{hyperref}
\usepackage{xspace}
\usepackage[numbers]{natbib}
\usepackage{balance}
\usepackage{xcolor}
\hypersetup{
  colorlinks=false,
  pdfborder={0 0 0}
}

\usepackage{fontawesome5}
\usepackage{framed}
\definecolor{formalshadelight}{RGB}{242,242,242}
\definecolor{formalshadedark}{RGB}{166,166,166}

\usepackage{csquotes}

\def\thepage{\arabic{page}}   % turn page numbers back on
\begin{document}

\title{All About That Bayes: A Focus Group on the State of Bayesian Analysis in Empirical Software Engineering}
% Alternative, more conservative title:
% \title{Updating Beliefs: A Focus Group on the State of Bayesian Analysis in Empirical Software Engineering}

\subtitle{ACM SIGSOFT SEN-ESE Column}

\numberofauthors{2} % number of author blocks
\author{
Justus Bogner\\
       \affaddr{Vrije Universiteit Amsterdam}\\
       \affaddr{The Netherlands}\\
       \email{j.bogner@vu.nl}
\and
Roberto Verdecchia\\
       \affaddr{University of Florence}\\
       \affaddr{Italy}\\
       \email{roberto.verdecchia@unifi.it}
}

\maketitle

\begin{abstract}
Frequentist statistics, most notably statistical hypothesis testing, are very well established for quantitative analysis in empirical software engineering (ESE) research, and arguably still its default approach.
However, Bayesian analysis as a promising alternative with several potential benefits is slowly but surely seeing more adoption.
As part of our regular column on empirical software engineering (ACM SIGSOFT SEN-ESE), we reflect on the state of Bayesian analysis with a focus group of experts: Julian Frattini, Richard Torkar, and Robert Feldt, all from the CSE department of Chalmers University of Technology and University of Gothenburg, who tirelessly educate the community about Bayesian analysis with PhD-level courses and conference tutorials.
Among other things, we discuss what advantages Bayesian methods hold for ESE research, how adoption and reviewer reactions evolved over time, common challenges faced while applying it, and how we as a community can make its adoption easier.
Finally, our experts recommend suitable materials for interested readers to get started.
The content of this paper is a faithful account of our conversation from November 6, 2026, which we moderated and edited for our~column.
\end{abstract}

\section{Focus Group Discussion}

\textbf{Moderators: We kindly ask everybody to briefly introduce themselves. Who are you, and what is your connection to Bayesian methods? For example, how did you come across them, or what made you stick with them?}

\textbf{JF:} Hi, my name is Julian Frattini. I'm a postdoctoral researcher here at Chalmers for 1 1\slash 2 years now. My connection to Bayesian methods, including the reason why I got into it and why I stuck with it, is actually in this meeting. During my PhD at the Blekinge Institute of Technology, Richard was offering a PhD-level course on this material, I think in 2021 or 2022. This was the single most transformative course that I had during my PhD. It really changed the way I did my research but also how I looked at the world, at least through the quantitative lens. It gave me a lot of drive during my PhD studies. I really felt like this is something that can have an impact. And that's also the reason why I am currently here at Chalmers with Richard and Robert because this is exactly the type of topic, the type of knowledge and capability that I want to further develop.

\textbf{RT:} I'm Richard Torkar, a professor in software engineering, and head of department for Computer Science and Engineering here at Chalmers and University of Gothenburg. I was always more interested in quantitative analysis. While I was a PhD student, I got a tip from my supervisor at that time to read a book, which I did. It's called Probability Theory: The Logic of Science~\cite{Jaynes2003}, and that got me really thinking. But at that time---this was maybe 2005---we didn't have the software, and we didn't really have the computational power in our labs as we have today. So, it was a bit early, but I was really enticed and followed it quite a lot in the coming years. And of course, the supervisor I had at the time was Robert. So, let's hand over to Robert now. \emph{(laughs)}

\textbf{RF:} It sounds like I'm the source of all this, I don't know. \emph{(laughs)} Yeah, I actually found the mentioned book by E. T. Jaynes~\cite{Jaynes2003}. I had a general interest in probability theory and information theory. During my PhD, this book was not yet published because he wrote it and made a PDF available, but he sadly died before he finished it. I found this PDF at the end of the 90s and started reading it. At the time, I was using information theory for software engineering problems, mostly for software testing. Jaynes derives probability theory from a small set of basic requirements for reasoning under uncertainty, and Bayesian inference follows directly from them; it's that fundamental.

Regarding my background, I did my PhD here at Chalmers on what we would now call search-based software testing. After finishing my PhD and moving to a different university, I became one of the supervisors of Richard. I think none of us had a plan to stay in academia, but we were recruited by Claes Wohlin. So then, we continued working together from~2006.

\textbf{Moderators: Okay, perfect, let's jump into the first question. It's probably fair to say that most ESE researchers have never run a Bayesian analysis for one of their studies. In two to three sentences, how would you explain what Bayesian data analysis is and what makes it different?}

\textbf{JF:} I think there are two aspects I always try to get across with people. First of all, what is it exactly? It's all based on Bayes' theorem. The general equation here is that you formulate \emph{priors}, i.e., prior assumptions about how the world works, and then use \emph{observations}, i.e., empirical data, to update those priors. This gives you a very principled approach to iteratively improving your knowledge about hypotheses and phenomena. Now, if we want to contrast this to the opposite camp called the frequentist approach, the interesting point is this: while the frequentist hypothesis testing estimates the probability of observing data at least as extreme as the one collected, assuming that a certain hypothesis is true, the Bayesian approach does the opposite thing. It estimates the probability of a hypothesis being true given certain observed data. This distinction is a bit counterintuitive at first. But once you relate it to the kind of questions that we're asking in ESE research or any empirical research, you will see that the Bayesian approach actually provides the answers to the questions that we're asking. We're less interested in how likely data is given a hypothesis. We want to know how likely it is that a hypothesis is true. And Bayesian methods give us precisely those answers.

\textbf{RT:} To add to this, if you look at the basic constructs here, we're talking about probability that describes our uncertainty about something we're estimating. When we update that uncertainty with data, we get a full posterior distribution and not a simple yes-or-no verdict. I think that's the important part. We're not trying to be black and white; we're trying to look at the nuances, the gray areas, and the uncertainty to get a better grasp of that. Additionally, in software engineering, we very often deal with the type of data that is very suitable to model with the type of approaches we use. That is to say, multi-level data, e.g., developers working within projects within companies, often based on repeated measurements and also in combination with small sample sizes. It gets quite messy. But if you use these types of approaches that Julian and I teach, then you deal with multi-level models, partial pooling, and other techniques that handle this very naturally and gracefully. So, in the end, as Julian was alluding to, you reach an answer to the question of what the probability is that an effect is bigger than $X$. That is very often what practitioners actually want to know. Because then, you can take that and derive the costs or time saved given this. But an honest caveat to this is, of course, that, with large sample sizes and simpler designs, then the two approaches give very much the same answers, whether they are Bayesian or frequentist, right? So, the real gain is explicit assumptions and better models.

\textbf{RF:} I would like to add one more big benefit here that is maybe a little harder to explain because it relates to the whole scientific field practicing this, not an individual study. But I think it's important to mention. What I consider to be a big problem, which I also mentioned a few years back in my keynote at ESEM 2019\footnote{\url{https://eseiw2019.com/keynotes}}, is that the SE community has something called neophilia. In computer science in general and especially in SE, every study needs to be novel. This means that all the studies we have are isolated islands, which is very hard to build on. When we build on something, it's in a qualitative and very unclear sense. Maybe I cite Roberto's paper, and I say (hypothetical example) \enquote{As Verdecchia et al. showed, larger team size leads to lower productivity.} That's a very high-level sentence, but I then build my new study based on this very crude summary of what Verdecchia et al. found. Maybe they found an actual probabilistic distribution of how team size affects productivity, right? Then, Roberto's posterior is the full output distribution that Richard was talking about. I can use that directly as my prior in follow-up research, i.e., I can collect new data and refine the knowledge that already exists. It's an explicit way to create a body of knowledge, for which we currently have a very vague and coarse-grained view. I think, scientifically, this is a big benefit of Bayesianism, regardless of all the machinery and the individual study benefits that we might see.

\textbf{Moderators: Thank you very much! Are there any other important reasons why we should choose Bayesian approaches over frequentist ones in ESE? I think Richard hinted at small sample sizes.}

\textbf{RT:} Yes, if you are careful about how you set your priors, e.g., with predictive checks, you can get away with smaller sample sizes. Those are unfortunately a curse in ESE research, like in many other research areas.

\textbf{JF:} I think there's a nice contrast between the two approaches here. A Bayesian model with well-justified priors trained on a small sample size will give you a more uncertain answer, but it's still a correct answer conditional on the model, whereas a frequentist significance test is more likely to give you a misleading answer altogether because it's more prone to error types (exaggerated or sign-flipped estimates) that regularizing priors guard against. One more thing to emphasize here: Richard already alluded to it, but Bayesian methods force you to be much more explicit. In frequentist approaches, it's very much a black-box approach. Basically, you follow a certain flowchart, make a couple of decisions about your data, and choose some model type. In Bayesian methods, you have to make explicit assumptions that you document and defend in your manuscript. That makes for a much more transparent scientific process for deriving conclusions from data, which is a great benefit.

\textbf{RF:} Yes, as a reviewer, you can now get into a meaningful discussion. Do I agree with the explicitly stated assumptions? Well, if someone uses a t-test, this also implies many assumptions that a big chunk of SE researchers probably will not even know about, but they are hidden. Bayesian analysis kind of forces you to describe them.

\textbf{Moderators: Are there also some cases, e.g., specific types of studies, where you would recommend going for a frequentist approach instead of going the Bayesian route?}

\textbf{RF:} You are asking the wrong people. \emph{(laughs)}

\textbf{RT:} To be honest, learning to use Bayesian statistics is also a cost. You have to invest time. So, making that investment because you suddenly face a situation where you feel that maybe you should do that is probably wrong. Sure, if you have a large sample size and a simple design, why not just do frequentist statistics instead, right? If it's the other way around, well, then Bayesian statistics is usually a bit more nuanced, a bit more balanced, and a bit more explicit in the choices we make, but it's also an upfront cost.

\textbf{RF:} Yeah, it's definitely more costly in learning and in some sense in getting it right. There are more knobs to configure. Actually, we should be open here. Even many of our published papers in the area of Bayesian analysis have rarely found a big difference in the frequentist and Bayesian analysis results. It rarely depends on that. So, I think you should not only do it for practical reasons. You should also think about what kind of researcher you want to be and what a more principled way of doing this could be.

\textbf{RT:} We should always do better if we can, and we usually can. Science is about trying to constantly strive for betterment in one way or another, isn't it? Either for the science itself or for society. So, doing better is a good rule of thumb that we should adhere to.

\textbf{RF:} Yes, this is not a clear answer, but it helps contextualize it. For the opposite, i.e., where it is strongly advised, it's not only small sample sizes and that you want to be more principled: our TSE paper actually shows how to leverage Bayesian methods to clearly show the practical significance of results for industry~\cite{Torkar2022}. There's a quantitative way of taking the output of a scientific study---the full distribution---to transfer this as a cost indicator for industry. This helps you to actually decide in a systematic way. If you have a frequentist approach, you will have to make a qualitative argument for why you should do something because there is no equivalent principled way.

\textbf{JF:} Yes, beyond the idealistic argument based on principles, I also think that the two main criteria are sample size and model complexity. And as Richard already said, we usually deal with small samples in SE. When it comes to model complexity, well, if we're only asking simple questions, then what are we doing as researchers? If we are actually striving to address hard problems, then we will meet our limits with these frequentist approaches, particularly when it comes to these kinds of clustered data or nested data, which is very common with repeated measurements. That's where Bayesian methods will leave frequentist methods behind very quickly.

\textbf{Moderators: Thank you very much! Can you share with us a success story, e.g., a study of yours or by someone else, where the Bayesian approach changed the conclusions or made them much more useful?}

\textbf{RT:} This is a bit harder because, as Robert mentioned, if everything is as it should be, then there shouldn't really be a big difference. But sure, we've had cases where we've seen differences. To what extent this depended on frequentist vs.\ Bayesian statistics is a bit hard to tell. I would say it rather depends more on knowing the basics of causal modeling. Many people don't understand causal modeling at all, and they simply think that you should take all the data you have---big data---and that this would be better data. But big data is \textbf{not} automatically better data. Big data can be very bad data. \emph{(laughs)} When I lecture, I say that I'd rather have 25 good data points than 25 million bad ones. It's the quality of the data that decides. When you have the data, deciding which of this data should be included in your model is more a causal question than a Bayesian or frequentist one. I tend to use Bayesian analysis because it's explicit, because it gives me a probability distribution instead of a yes-or-no answer. But sure, we've seen a few cases where it definitely makes a difference, very often when it comes to smaller sample sizes.

\textbf{RF:} I agree with Richard. It's usually not like everything changed in a study when we switched to Bayesian analysis. However, there is one other thing. We have seen that we get a more fine-grained understanding with Bayesian methods. We wouldn't even have posed certain questions with frequentist methods because it would have been unclear how to study that with the limited toolbox. It can be done, but you will have to be much more of an expert to put together a frequentist method for such complex situations. Especially for some of our studies with Carlo Furia, we got a more nuanced result due to Bayesian methods. So, it is more this richer understanding rather than totally different conclusions. This, to be honest, I'm not sure I can remember seeing.

\textbf{RT:} Like Robert said, this nuanced perspective can also help us question previous results, e.g., because they have been explicitly stated as something like \enquote{$X$ causes $Y$}. When we do a Bayesian analysis of it, we very often come to the conclusion that there is no clear effect. There might be a small effect, but it's really not such a strong one. And if you look into the effect distribution, you see that it's maybe not such sexy results as previously thought.

\textbf{JF:} Actually, we have a study where we did exactly this~\cite{frattini2024second}. We took a paper and went through multiple steps of revising the data analysis in one step, replacing the frequentist with a Bayesian method. The Bayesian analysis was more complex but also more true to the data-generating process and also revised the causal assumptions. The latter was usually the most significant change in the results, where you really saw previously assumed effects completely disappearing. For the usual Bayesian analyses that we do for our studies, there is not really a frequentist counterpart to compare it with. So, it's hard to clearly answer that question, though it would be interesting. I must say that my fingers are itching to make a demonstration out of that. \emph{(laughs)}

\textbf{RF:} Just to add to that, I think my view also changed a bit over the years. I think around 2017 and later, both Richard and I were really trying to preach that everyone should go towards Bayesian methods. But today, I think convincing someone to use causal analysis is much more important than switching from frequentist to Bayesian methods. On the other hand, the number of questions you can answer if you go for causal analysis with frequentist methods will be quite limited. Bayesian and causal analysis go hand in hand because Bayesian methods allow you to model more realistic and more complex situations. Causal analysis is more important, but they do strengthen each other.

\textbf{JF:} There's a great post by Richard McElreath\footnote{\url{https://x.com/rlmcelreath/status/1611391974485164034}}, who's a very prominent figure in this space in both Bayesian data analysis and statistical causal inference. Back in the day when Twitter was still a thing, he put this very nicely in a tweet with pictures, as he is typically also cited as \textbf{the} person to advocate for Bayesianism in data analysis. But even he is very explicit about causal inference being the real deal.

\textbf{Moderators: Maybe that should be the next focus group then at some point. \emph{(laughs)}}

\textbf{RF:} Absolutely!

\textbf{Moderators: All right, then let's switch to the current state of Bayesian methods in ESE research. Has adoption grown noticeably in recent years, is it still a niche, or something in between?}

\textbf{JF:} In my perception, it still remains a niche. As of right now, other than statistical causal inference methods, Bayesian data analysis is still adopted in a very limited fashion. If you see a published SE paper using Bayesian data analysis, you can typically easily guess who the authors are. And it's likely that one of us is behind it. But I think, for about 80\% of the papers that we submit with Bayesian data analysis, we get a very positive response from reviewers. Typically, we get reviewers expressing that they appreciate the level of detail in the data analysis. Of course, we take extra care to explain it. In 20\% of cases, you'll still have reviewers that say I don't see why this complexity would be warranted to analyze that data. So, it has not taken off to the degree that we would like to see yet, but we see in adjacent fields that this becomes more and more adopted. I think it was either this or last year that the US Food and Drug Administration finally adopted Bayesian methods for their approvals.

\textbf{RF:} Also in medicine. I do quite a lot of studies on medical AI, and medical researchers are, for good reasons, much more cautious about new methods than we are; in a sense, they have the opposite of neophilia. When patients' health is at stake, you want to be sure that a method works before you trust it. So, for a long time, the established frequentist methods, in which they are very skilled, have been the default. Now that the FDA has issued draft guidance on using Bayesian methods in drug and biologics trials, this could be a big driver for adoption in the coming years. But yeah, if we take our TSE paper~\cite{Furia2021} by Carlo, Richard, and me as a proxy for adoption, it now has over 100 citations on Google Scholar. It started slowly, but it is now cited steadily every year. So, there seems to be some objective evidence that Bayesian analysis is at least used more than before. But it's for sure been slower than we expected; let's put it like that. We had an ICSE tutorial in 2021\footnote{\url{https://conf.researchr.org/track/icse-2021/icse-2021-Technical-Briefings\#Bayesian}}, which was very well received. I don't know; maybe COVID killed the initial momentum. \emph{(laughs)} Because we had a lot of people in that tutorial, and then we did not see as much effect as we were hoping for.

\textbf{Moderators: When you read a Bayesian ESE paper today, do you think that the method section is typically described well? Or do you sometimes also see something like \enquote{frequentist thinking in Bayesian clothing}?}

\textbf{RT:} Actually, you have a good point there. I would say that the majority of papers with a Bayesian analysis that I reviewed do a tremendously nice job, in my humble opinion, I have to add. But you have a point that, in some way, they often want to replace the frequentist notion that the $p$-value is below $0.05$. But then they start talking about how the $95$\% interval excludes $0$ and so on. They still want to keep that type of black-and-white thinking because I think that they assume that reviewers are used to that storytelling and want to hear it that way. But I completely understand where they're coming from.

\textbf{RF:}  Yeah, and as with medicine, I don't think this caution is unique to Bayesian methods. Even when we have used LLMs to support the analysis of free-text data, such analyses tend to end up in supplementary material rather than in the main paper. That caution is understandable, since new methods need to earn trust, but it shows that there is researcher inertia to overcome in any field. You have to be both interested and curious to want to explore this more, so it will be a learning curve. How many SE research labs do we know that actively use Bayesian analysis? Maybe 5, maybe 8, but probably not many more than 10. I may well be wrong, though.

\textbf{JF:} No, I believe so as well. And funnily enough, for the story that Richard just told, I'm one of those people. \emph{(laughs)} Whenever we report a Bayesian data analysis in our papers, we typically first report the results as we would for someone who knows their Bayesian stuff. And then, it's almost literally a disclaimer: \enquote{If you are a Bayesianist, stop reading here. And if you don't fully understand what we've explained so far, then here's the one-line summary, the yes-no answer.} There are also methods for this, like \emph{region of practical equivalence}, which boils down the results into a categorical or even binary answer. But, of course, that's not what we're aiming for, even though this is still consistent with the analysis. At the moment, however, this kind of language interface still feels necessary.

\textbf{Moderators: Thank you. Very interesting. I think this is perfect for switching the topic to reviewing. How do reviewers generally react to your Bayesian papers, and did you notice that this changed over the years?}

\textbf{RT:} I can only speak for myself. It was extremely painful to publish because very few people had any idea of what we were doing. Every time we published something, we needed a replication package with the analysis in detail to have ample room for explanations. Still, every editor said we should spend a lot of time explaining our methods. There was a lot of repetition in the paper to explain what we're doing. My feeling is that things have changed now. Editors and reviewers accept the methods. If they understand it, I don't know, but they accept it in a different way compared to what they did before.

\textbf{RF:} Yes, that is a big change. We always had to add 2 to 4 tutorial pages to every paper before. I don't think there is a need for that now.

\textbf{JF:} I would say publishing a paper with a Bayesian data analysis is something completely different from publishing a regular frequentist analysis paper. Thanks to the groundwork of Richard and Robert, I am in the situation where I have received a lot of positive responses, roughly 80\% positive compared to 20\% negative sentiment towards the Bayesian part in reviews. Bayesian methods get now accepted at major software engineering conferences. This year we managed to get very positive responses both from ASE and ICSE, even from the main tracks. But you have to structure a Bayesian paper completely differently. This alone is one of the major challenges of adopting Bayesian methods. In a frequentist analysis, there is a mutual understanding of what people are doing when they describe it, e.g., the typical \enquote{We check for normal distribution and then use this test.} There is broad agreement that this is nearly everything you need to write. In a paper with Bayesian data analysis, the best that we could do so far was 1 or 2 pages of IEEE double-column. That is the least amount of space we need to report our data analysis. That is a commitment, especially for big, dense papers, where you need a lot of space. The Bayesian data analysis sections are intended for someone who understands them and can follow all our explicit decisions, such as our priors or the probability distribution that we used for describing the outcome variable. Our pragmatic intention right now is that we need to explain our approach enough that a reviewer who does not know Bayesian data analysis can still follow or at least believes us, and those are two completely different things. I hope one day Bayesian methods will become commonly accepted knowledge, but right now, there is still the need for this tutorial style to describe our approach. We have to teach the reviewer something, which I think is also a nice thing; at least I enjoy learning something new when reading a paper. But this tutorial style turns the review around. The reviewer should be the expert. Right now, we need to give reviewers enough information so they can follow what we are doing, but they will not be able to pick out the small details, such as using the wrong distribution. That is the kind of conversation that we would like to have through reviews, and we hope we will get there.

\textbf{RF:} I do not want it to sound negative for some readers, but one risk with Bayesian methods is maybe that you write more explicitly what you did. This also can make it easier for a reviewer to have something to critique. It would probably also be a critique for a traditional analysis, but it is not so explicit, so there's less of an attack surface.

\textbf{RT:} To get back to the question, I would say, yes, we notice a difference compared to the past. It is better now. We don't get any more comments such as \enquote{Priors are subjective, full stop. I reject the paper.} Or \enquote{Where are the $p$-values? I reject the paper.} So at least, we don't see such comments.

\textbf{RF:} Oh, yes, exactly. It's accepted enough that you will not get these, excuse my wording, uninitiated comments.

\textbf{Moderators: Speaking of uninitiated comments: Do you have any advice for a reviewer with no background in Bayesian statistics who has to review a Bayesian paper?}

\textbf{JF:} Tough question. They should focus on a couple of things. First off, at the core of any Bayesian analysis is a regression model, and a lot of knowledge from frequentist analysis can be transferred for the structure of that regression model: what predictors are mentioned, how are they related to each other, and so on. That is something arguably more related to causal inference, but it is an important part to look out for. Priors are usually recorded, and there is usually very little to critique since ESE as a field is still so young. There is very little strong evidence about empirical effects. So, most priors are usually very conservative and very uninformative to not bias the analysis too much. Whenever you see strong prior assumptions as a reviewer, you should really check whether the authors provide sufficient support for this.

Another important part of a Bayesian analysis is whether authors report their diagnostics. Bayesian methods give you a lot of insights into how reliable the obtained results are. There are a couple of metrics people can easily look into without needing to understand how they are produced. Clearly defined thresholds that need to be met. For more information on this, interested readers could look into divergent transitions and the so-called $\hat{R}$ (\enquote{R-hat}) value. If you see these metrics in a paper, you can typically compare them to some commonly defined thresholds, and if they are met, that at least speaks for the stability and the reliability of the model. As a reviewer not familiar with Bayesian analysis, these are the two top things I would analyze to still produce a relevant review.

\textbf{RF:} I would like to add one thing. I think what would help the community, and in particular the reviewer not familiar with Bayesian analysis, is if we wrote a high-level supplement to the ACM SIGSOFT Empirical Standards for Software Engineering\footnote{\url{https://www2.sigsoft.org/EmpiricalStandards}} to describe how you judge a Bayesian analysis. With that supplement, you will not become an expert, but at least we will be able to build a shared consensus among the groups specializing in Bayesian analysis to alleviate some of the review barriers.

\textbf{RT:} Just to clarify, using Bayesian inference is a way to estimate the unknowns in a probability model, any probability model. It does not have to be a linear model. It can be non-linear, but it can also be very simple cases where you are not in need of a regression analysis at all. It is really up to what you need, what type of data you have, what the problem looks like, and what the design of the study is. But very often we indeed fall back to different types of linear regression models, because they are powerful and common.

\textbf{Moderators: You hinted at this before. In your opinion, what are the biggest challenges Bayesian researchers face?}

\textbf{RT:} I see people struggling with misspecified models. Gelman calls it \enquote{The folk theorem of statistical computing}.\footnote{\url{https://statmodeling.stat.columbia.edu/2008/05/13/the_folk_theore/}} If you get problems, it is usually because you are misunderstanding something. For example, you have specified a model in a way because you assume the underlying data generation process gives you certain data, but instead you made the wrong assumption about the data generation process itself. Whatever algorithm you use will not be able to solve the problem for you. The multi-dimensional space that it's searching contains geometries that are painful for a sampler. In practice, an \enquote{unsolvable problem}. And then you get these very funky errors. Every time I hit something like that, it has been because I misunderstood something. My assumptions about the underlying process~have~been~wrong.

\textbf{JF:} When it comes to computational cost, nowadays it is possible to run a Bayesian analysis on any regular laptop. It is still more expensive than frequentist analyses, though. A Bayesian model would fit maybe in minutes, which is a lot more compared to a t-test usually taking milliseconds. People would need to recalibrate their expectations a bit. But I think the most difficult entry barrier is how difficult it is to set up the environment. Most of the analyses we run use a Markov Chain Monte Carlo algorithm, which is a brilliant algorithm, but you need a specific environment for that. One of the best environments is implemented in a language called Stan, which requires its own interface and environment. It's not rocket science, but you will spend an hour or two on setting this up. It is much more cumbersome than just importing a library that has several tests pre-implemented. Perhaps if it was easier, more people would give Bayesian analysis~a~try.

\textbf{RF:} I think the main barrier a Bayesian researcher faces is a reviewer that is less familiar with these methods. For the general software engineering researcher, it's the learning barrier of how and where to start: why should they need to spend time learning something new when they can just use what is already accepted. All technical problems are in my opinion far less important.

\textbf{Moderators: Changing the topic, which would be your main argument to convince a frequentist researcher to use Bayesian methods?}

\textbf{RT:} You get a richer understanding of what you are studying. A better grasp of uncertainty, which propagates through the models, to be able to paint a better picture of where we are and where we are heading.

\textbf{JF:} The idealist in me would try to convince people that this is simply the right thing to do. It gives you the right answers to the important questions. Richard once presented it very nicely: \enquote{Bayesian analysis lets the data speak its language.} And all without enforcing things, without pressing data through black boxes to just get out a very coarse binary label at the end. Bayesian analysis really lets you understand the data and its intricacies. It is the tool we will ultimately need to answer the real complex questions. And I think this is where we should be headed, as simple questions will soon be solved automatically with ease.

\textbf{RT:} I think, as alluded to earlier by Robert, an important point of Bayesian analysis is that you get a posterior out of it, which you can then use as a prior. This means that the precision of our analysis will be better: the uncertainty will decrease over time when we get more evidence. Don't get me started with measurement errors, which most software engineers do not even think about, because that makes it hard for me to sleep at night. We have completely missed it there. Measurement errors are basically the only thing the physicists are obsessing about, but in our software engineering world, they don't exist.

\textbf{Moderators: How do you think Bayesian methods position themselves within the ESE research field? What needs to happen for them to become a normal part of the ESE toolbox?}

\textbf{JF:} Lowering the technical entry barrier is definitely one aspect of it, but the big thing will be disseminating Bayesian analysis knowledge. We need a cultural shift. Everybody knows frequentist methods, and everybody expects frequentist methods, so we are stuck in this situation. There is too much disciplinary inertia to move away from frequentist analyses. We realized that we need to start this dissemination of Bayesian analysis knowledge, e.g., in the form of conference tutorials. Richard and I are also now working on a PhD course for students of different universities. There are great resources and textbooks to learn from. It is finally accessible also on your own machine, which was not the case 20 years ago. We think it is the best time to learn Bayesian methods, but it is really helpful to have someone to guide you through it. Compared to frequentist analysis, with the Bayesian one you have to be much more conscious about what you are doing, much more explicit, much more transparent. And for that, you need to feel secure enough to present a study based on that. We believe education is the number one solution.

\textbf{RF:} I want to add something that might convince more people to look into Bayesian methods. Everything we have discussed up until now is about the meta level. We are only discussing Bayesian methods as a tool in the data analysis of our studies. But the benefits of using Bayesian methods carry over to software engineering practice and technologies. For example, instead of viewing code execution in binary terms, like whether a branch is covered or not, you could take a probabilistic view of it and build Bayesian methods into our tools and techniques. There are quite a few papers doing this in software engineering over the last 10 years. Maybe that is an entry point, too. If we can show that you can actually build better software technologies that benefit software engineering researchers and practitioners, that might be a gateway to broader use. I have worked on this with Shin Yoo's group at KAIST, and quite a few groups have now done work in this direction, I would say more than there are people using Bayesian methods to analyse their empirical software engineering data.

\textbf{Moderators: As a concluding last question: what is a concrete recommendation for readers who want to start tomorrow? Maybe a textbook, a favorite paper of yours, a tutorial, etc.}

\textbf{RT:} \enquote{Statistical Rethinking} by Richard McElreath, the 2nd edition~\cite{McElreath2020}. I know he is working on a third edition as well. I use it in our course, and we also use it as a textbook for our course for PhD students up to professors. It's an excellent introductory text because it focuses on skills. It focuses on you doing things and not only reading them. Another book called \enquote{Bayesian Workflow}~\cite{Gelman2026} was published very recently, which is a good book because working with Bayesian research is really a workflow and not a test. We build, we design, we check, we revise, we redo, and we iterate on this workflow all the time. I also think that book is excellent, with lots of good examples of how you can do Bayesian research. Those two books are a great start.

\textbf{JF:} Absolutely, \enquote{Statistical Rethinking} is the key reference people should pick up if they are serious about Bayesian analysis. I would add one more resource that Robert and Richard authored, \enquote{Bayesian Data Analysis in Empirical Software Engineering Research} by Carlo A. Furia, Robert Feldt, and Richard Torkar~\cite{Furia2021}. That should be the first reference that people pick up. Statistical Rethinking is fantastic, but it is a commitment. If you really want to dig into it, you will have to reserve a lot of time, and you will likely not fully comprehend it with the first reading. While the paper of Furia et al. can be read in an afternoon, it is much more concise and very focused on Bayesian data analysis.

\textbf{RT:} And then, of course, during spring 2027, Julian and I will again run a course, and we will open it up to the world this time.

\textbf{Moderators: Thank you so much for your time and insights! This was really interesting.}

\bigskip

\section{Conclusion}
Looking back at our conversation, we have the lingering feeling that we talked about more than just Bayesian analysis. In the end, we might have touched upon something far deeper, namely how frequentist research approaches established themselves as mainstream in software engineering due to a mix of heterogeneous and at times seemingly random factors, such as past computational power, learning curves, existing tools, and peer-review knowledge and preconceptions. Taken together, perhaps these factors made the frequentist route the easiest one, to the point that we nowadays rarely question it. Yet, as empirical researchers, we know that science benefits from looking at the same phenomenon from different perspectives. As our experts noted, the two approaches usually agree, but Bayesian analysis adds explicit assumptions, richer models, and a natural way to build on prior evidence. Perhaps, as a research community, we should resist the pull towards what is simpler, more widely accepted, and easier to defend, and instead allow our research to mature from a methodological standpoint as well. For this reason, a growing adoption of Bayesian methods in the ESE community seems---in our opinion---desirable. As a broader takeaway, no analysis method should be adopted (or dismissed) just because \enquote{everybody is using it}. Choosing our empirical tools consciously, adapting them, and being able to clearly motivate such choices, is what really makes empirical research worth building~upon. 

\section*{Acknowledgements}
We kindly thank our three experts. We genuinely cherished their valuable time and insights!

\bibliographystyle{plainnat}
\bibliography{biblio}

@book{Jaynes2003, place={Cambridge}, title={Probability Theory: The Logic of Science}, publisher={Cambridge University Press}, author={Jaynes, E. T.}, year={2003}}

@ARTICLE{Torkar2022,
  author={Torkar, Richard and Furia, Carlo A. and Feldt, Robert and de Oliveira Neto, Francisco Gomes and Gren, Lucas and Lenberg, Per and Ernst, Neil A.},
  journal={IEEE Transactions on Software Engineering}, 
  title={A Method to Assess and Argue for Practical Significance in Software Engineering}, 
  year={2022},
  volume={48},
  number={6},
  pages={2053-2065},
  doi={10.1109/TSE.2020.3048991}}

@ARTICLE{Furia2021,
  author={Furia, Carlo A. and Feldt, Robert and Torkar, Richard},
  journal={IEEE Transactions on Software Engineering}, 
  title={Bayesian Data Analysis in Empirical Software Engineering Research}, 
  year={2021},
  volume={47},
  number={9},
  pages={1786-1810},
  doi={10.1109/TSE.2019.2935974}}

@book{McElreath2020,
	address = {Boca Raton},
	edition = {Second edition},
	series = {Chapman \& {Hall}/{CRC} texts in statistical science series},
	title = {Statistical rethinking: a {Bayesian} course with examples in {R} and {Stan}},
	isbn = {9780429029608 9780429639142},
	shorttitle = {Statistical rethinking},
	publisher = {CRC Press},
	author = {McElreath, Richard},
	year = {2020},
}

@book{Gelman2026,
author = {Gelman, Andrew and Vehtari, Aki and McElreath, Richard and Simpson, Daniel and Margossian, Charles C. and Yao, Yuling and Kennedy, Lauren and Gabry, Jonah and B{\"u}rkner, Paul-Christian and Modr{\'a}k, Martin and Leos Barajas, Vianey},
title = {{B}ayesian Workflow},
publisher = {Chapman \& Hall/CRC},
address = {Boca Raton, FL},
year = {2026},
isbn = {9780367490188}
}

@inproceedings{frattini2024second,
  title={A Second Look at the Impact of Passive Voice Requirements on Domain Modeling: Bayesian Reanalysis of an Experiment},
  author={Frattini, Julian and Fucci, Davide and Torkar, Richard and Mendez, Daniel},
  year={2024},
  booktitle={International Workshop on Methodological Issues with Empirical Studies in Software Engineering (WSESE'24)},
  doi={10.1145/3643664.3648211}
}

\end{document}